# A Conceptual Design of In-Game Real and Virtual Currency Tracker

Dennis Barzanoff

School of Computing and Communications, Lancaster University, 04109 Leipzig, Germany

d.rumenoffbarzanoff@lancaster.ac.uk

Amna Asif

School of Computing and Communications, Lancaster University, 04109 Leipzig, Germany

a.asif2@lancaster.ac.uk

The gaming industry is earning huge revenues from incorporating virtual currencies into the game design experience. Even if it is a useful approach for the game industry to boost up their earnings, the unidirectional and bidirectional in-game virtual currencies can invoke inadequate gaming behaviors and additions among players. The market lacks gaming and customer protection regulations to avoid the financial, behavioral, and psychological exploitation of users. Therefore, it is needed to develop visual or textual interface design recommendations that help the game players keep balance in their spending and improve their gaming behavior. This paper presents a conceptual design of an in-game purchasing module that allows the user to observe their real time spendings in relation to virtual currency buying.

## 1 INTRODUCTION

Virtual currencies are generally issued and controlled by software or game developers and accepted among specific visual communities such as game players. Virtual currencies can be divided into three types: (1) closed virtual currency, (2) virtual currency with a unidirectional flow, and (3) visual currency with a bidirectional flow. The closed virtual currency does not have any connection with real money. It means that closed virtual currency can only be earned by competing for some in-game activities. However, unidirectional and bidirectional currencies have a connection with real money. We can only buy unidirectional virtual currency with real money. But bidirectional virtual currency can be bought and sold using real money [1]. The game industry products and services, including in-game virtual currency purchases, are generating millions of dollars in revenue [8]. According to statists, a video game annual expenditure per customer in the year 2021 was 11.94 U.S. dollars [12]. Skins, which can be thought of as a type of in-game currency, are a prevalent means of betting in online games [11]. Every 1 in 10 children has tried skin gambling between 13 to 18 years old. A 13-year-old boy told the parent zone that he had lost 2000 British pounds of virtual currency on gambling sites [13]. Skins are attractive for kids because by being bidirectional it allows them to buy and sell their virtual currencies for real money. Besides skins, the other virtual currencies are gold, time token, simoleon, emerald, coins, renown, points, timber, metal, wealth, gem, elixir, cash, bucks, diamond, etc., [1]. Such currencies allow players to buy and sell in-game virtual properties, accelerating their level, and increasing game-winning probabilities by unveiling many cheat tricks etc. Microtransactions are low-cost expansions of existing games and lead to some hidden costs [2]. Microtransactions are comprised of many factors (such as buying in-game benefits or social aspects of the game) that encourage the players to spend on in-game items as it can be regarded as predatory monetisation [6]. Such schemas are designed to encourage players to spend more money than they initially intended, potentially leading to overspending or addiction. In-game purchases are a very advantageous and lucrative way for the gaming industry to generate huge revenues by making use of these trading tactics. Also, another reason in-game currencies are being used is because by being able to sell an item for a virtual currency, players can then use the currency to buy other items, which facilitates trading. However, disadvantages exist for the players in terms of uneven playing field between players who spend real money on virtual items and those who do not. This can lead to frustration and a feeling of unfairness for players who do not have the financial means to compete on the same level. Game developers have employed various techniques to exploit the reward systems in the brain, such as offering randomised rewards, creating social pressure to spend money, and implementing time-limited events. These strategies create a sense of urgency and excitement, leading users to make impulsive and sometimes excessive in-game purchases [9]. While the players are trapped with predatory monetisation, they cannot realise the amount spent in terms of real money as there is no way of tracking it. Similarly, there is no legal consensus related to the activities that fall under gambling or gaming [7] [4]. The use of psychological tactics in mobile game design has become a prevalent issue in recent years, as it has been linked to the rise of problematic gaming behaviours and addiction.

Previously many studies have been conducted to try to identify the issues and intentions of in-game currency purchasing. Evers et al., conducted an investigation in one survey and two experimental studies. They found that although microtransactions help players in their games, unfortunately players are less respected and not supported by the other players in the games [2] Marder et al., conducted an interview with 32 players to explore their motivation for buying virtual items within free to play games. They found that hedonic and social motivations are dominant [10]. Although players have social motivation, they are not as respected among the players as reported in [2]. Due to difficulty in the analysis of the cost and benefits of items, the notion of 'value of money' was rarely discussed. Therefore, the authors suggested that there is a need to inform the players regarding the value of money while buying virtual goods [10]. Fang has conducted a study to observe the impact of the exchange ratio of the home and in-game currencies. The author found that if the home currency is much more valuable than the in-game currency, players tend to spend more money because of the illusion that they are repurchasing more currency [3]. Hamari et al., reviewed existing research to investigate the player's in-game purchasing behaviour; they found that reasons for unobstructed play, social interaction, and economic rationale are directly related to the amount the players spend [5]. These studies reveal that games are designed to motivate them to spend money to upgrade their virtual social status and unlock to gain maximum game experience. However, the studies are limited to providing some possible solutions that can save players from the harmful consequences of in-game purchases.

It has been shown that many individuals have developed a dangerous disconnection between the money they spend in-game and their real-world financial resources. They may overspend on virtual goods without fully comprehending the monetary value of those transactions or the long-term impact on their finances. This issue is particularly prevalent among young people who are still developing their financial literacy and decision-making skills. Furthermore, the increasing accessibility and popularity of mobile gaming have exacerbated the problem, as it has become easier for people to spend money on games anytime and anywhere. Consequently, the potential for harm is significant, with some users facing financial distress, debt, and other negative consequences due to their in-game spending habits. King and Delfabbro highlighted the requirement of such payment options, which can display real currencies to the players. They suggested that loot boxes must always be visually related to their real-world costs, not just with virtual currencies [8]. In response to these concerns, it is crucial to develop new technologies and systems that promote responsible gaming behaviour and raise awareness about the impact of in-game transactions on real-life finances. Implementing the new awareness-oriented system designs will be a promising step towards achieving accountable gaming behaviour. By providing users with precise and real-time information about their in-game spending, the system can help individuals become more conscious of their actions and make informed decisions about their in-game purchases. Designing such a solution has come as the main challenge of drafting this paper. The approach taken in this paper is to identify specifics in the user experience that contribute to the lack of interpretation of in-game currencies.

## 2 MATERIAL AND METHOD

This section aims to explain the conceptual design of an in-game interface that is supposed to promote real-time user awareness of the mapping between physical and virtual currencies. The design is motivated by interviews, which were conducted with three regular game players. They responded that they pay for ad-free games experience, upgrades, obtaining extra resources, and for faster progression in the game. One of the players answered that the gambling effects victimised him. The results helped us determine that users have two serious problems with mobile games: (1) they are often used as a gambling profit tool by companies, and (2) users are losing track of how much money is going into the specific parts of the game with time.

In this regard, we proposed a concept of an additional system that be placed between the user and the game that can collect information and show the user warnings and specific information that ought to make them aware of their spending choices. The proposed conceptual system is presented as a prototype that combines the visual and written recommendation of in-game purchasing design. The prototype is developed based on three main requirements:

1) The design must show the user how much money they have spent on each virtual currency app.
2) The information needs to be categorised by dates.
3) The users will be presented with details about how the system induced the result.

One of the preconditions of the proposed system is that it should be integrated with the specific video game so it can gather the necessary data. This must be done by integrating every video game with the payment platform and requiring video game providers to report in-game purchase data to the Operating System (OS) provider. The information about in-game currency, data, and time purchase is input into the system. Comments on the legal ground for collecting such information are out of the scope of this paper.

The other important factor to consider is the manner in which users have spent a specific amount of money on the virtual product. First the player buys a virtual currency, then another currency with that currency, until they eventually obtain an item. One problem that can arise is when the game allows reselling of items. The difficulty is there can be an inflow of in-game currency from the game itself, from reselling something, which makes the calculation of the value of the original real world currency you spent more difficult. The page can only show buying's and ignore reselling's. To solve this problem simply, the proposed solution is to pick a strategy for using inventory, such as first in first out FIFO or last in first out LIFO. If the user buys an item, which would add an entry to the page, then they sell the item for new in-game currency, buy more in-game currency, use them together to buy something whose value is greater than both of them, the system should only capture the part that involves the money spent by the user. This should be easy to implement if the system has full access to the changes in the in-game currency account logs by using the equations (1) and (2).

Given that we have some base currency, such as dollar ($), we can calculate the exchange rate for the next currency and the next currency, until the user finally exchanges currency for an in-game item. Let a matrix E have k

columns and 1 row, containing the exchange rates, calculated based on prior indexing of the purchases, p is the price of the items, c is the number of items purchased:

$$E = [E_0, E_1, ..., E_{k-1}] \quad (1)$$
$$p = E_{k-1} \cdot E_{k-2} \cdot ... \cdot E_1 \cdot E_0 \quad (2)$$

## 2.4 Case Study

The user wants to purchase something in the game (fig 1(a)) but realises they don't have enough in-game resources to do that, so they resort to spending real money in the game. Then the user goes to the in-game shop to complete the purchase (figure 1(b)). After that, a window where they can complete the purchase, change payment method, address, etc. Now they can go into the settings page and, using this new window, they can observe their purchases and can see the conversion from real money, not into in-game currency, but into their items directly, enabling them to understand that they spent real money on an in-game item, increasing their awareness. This can be explained with the following scenario. The user wants to purchase a magic chest in the mobile game Clash Royale, As seen in Figure 1(a), and they only have 249 out of the necessary 250 gems. However, for the sake of simplicity in the calculation, let's assume they have 0 gems, not 249. They go into the "Gems" section to purchase more gems, but the user finds more lucrative offers and purchases more gems than they need for the item. Eventually they settle for buying $19.99 worth of Gems, which gives them 2,500 Gems. Then they purchase the magic chest for 250 Gems. The behind the scenes, the system, having obtained this information, a calculate can be made that the magic chest costed $\frac{250}{2500} \cdot 19.99 = 1.99$, which is the same as the value observed in figure 1(c).

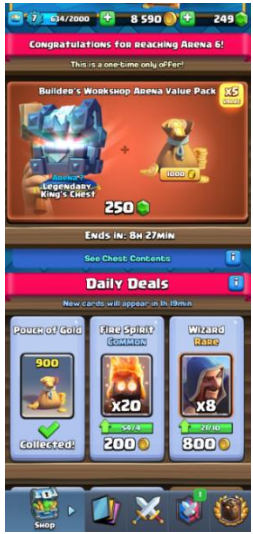
(a)

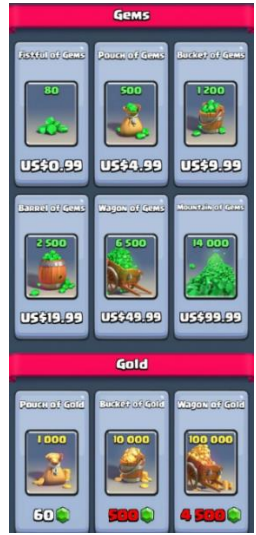
(b)

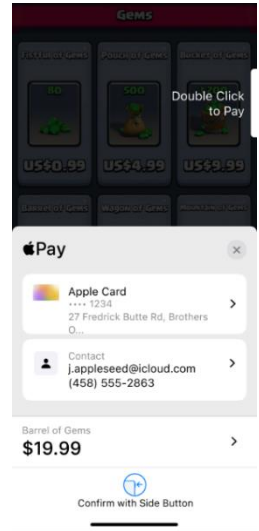
(c)

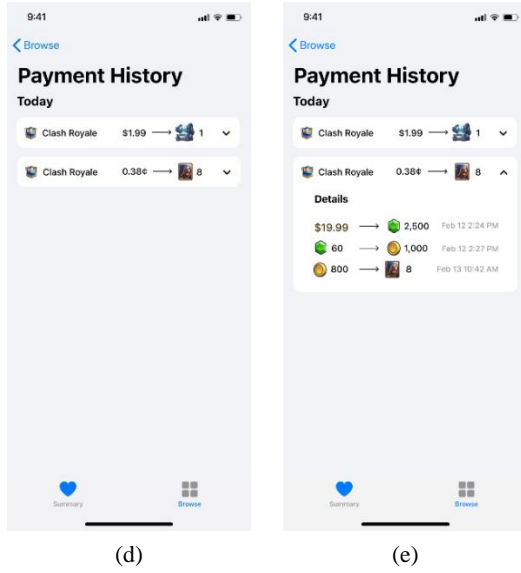

(d)            (e)

Figure 1: Flow of in-game purchasing: (a) User wants to buy something, but doesn't have enough in-game currency, (b)The users go to the shop to buy some. (c) User buys more gems than they need (d) After they make the purchase, they can review where their money went., (e) They could also see a more detailed view

Then the user decides to purchase 8 wizard cards from the store, which require another in-game currency, coins, used to buy cards in this game. However, coins are only sold in packs, for example, if we assume the user has 0 gold, they would need to buy a pack of 1,000 Gold. If that's the flow, the user then has 1,000 gold, they can proceed to buy 8 wizard cards for 800 Gold. When that happens, the system can calculate how much money was spent on wizard cards. We can convert from wizards to coins, from coins to gems, and from gems to dollars to estimate how much the wizards costed in real money.

First, we need to convert from the item (the 8 wizards) to gold coins. The exchange rate is the wizard to gold conversion. In this case we can convert 8 items to gold: $E_2 = 1 \cdot 800$. Then we need to calculate the conversion rate from gold to gems: $E_1 = \frac{800}{1000} \cdot 60 = 48$ . Then we need to calculate the conversion from this gold to gems:

$$E_2 = \frac{19.99}{2500} = 0.0079$$

$$p = E_1 \cdot E_2 \cdot E_3 = \frac{800}{1000} \cdot 60 \cdot 0.0079 = 0.38 \text{ ¢}$$

In the end we find that the wizard items costed 0.38 ¢, as shown in figure 1(c) and 1(d)

## 3 CONCLUSION

In conclusion, our study has shown that the concept of a new system design that has the potential to increase self-awareness among individuals who engage in in-game currency transactions. The prototype presented the real-time information about the conversion from in-game currency to real currency. It can help users to understand the true value of their virtual assets and the potential costs of their gaming habits. We believe that this proposed idea can be a valuable step in promoting responsible gaming behavior and reducing the risks of gaming addiction. This work is in progress and required to be evaluated. Therefore, further research is needed to fully evaluate the effectiveness of this approach and to identify any potential drawbacks or limitations of the system.